\documentclass[conference]{IEEEtran}
\IEEEoverridecommandlockouts
\usepackage{cite}
\usepackage{amsmath,amssymb,amsfonts}
\usepackage{algorithmic}
\usepackage{graphicx}
\usepackage{pifont}
\usepackage{multirow}
\usepackage{textcomp}
\usepackage{url}
\usepackage{xcolor}
\def\BibTeX{{\rm B\kern-.05em{\sc i\kern-.025em b}\kern-.08em
    T\kern-.1667em\lower.7ex\hbox{E}\kern-.125emX}}
\begin{document}

\title{Structural Inference in Undocumented Mobile Databases:
A Reproducible Benchmark for Evaluating Agentic Reasoning in Digital Forensics}

\author{\IEEEauthorblockN{Jeel Piyushkumar Khatiwala}
\IEEEauthorblockA{\textit{School of Criminal Justice} \\
\textit{College of Public Affairs} \\
\textit{University of Baltimore} \\
Maryland, USA \\
Jeel.khatiwala@ubalt.edu \\
0009-0001-3059-9339}
\and
\IEEEauthorblockN{Divyangkumar Patel}
\IEEEauthorblockA{\textit{Consumer \& Marketing Solutions SRE Team} \\
\textit{Cox Automotive Inc.} \\
Atlanta, GA, USA \\
Divyangpatel38@gmail.com}
\and
\IEEEauthorblockN{Weifeng Xu}
\IEEEauthorblockA{\textit{School of Criminal Justice} \\
\textit{College of Public Affairs} \\
\textit{University of Baltimore} \\
Maryland, USA \\
wxu@ubalt.edu}
}

\maketitle

\begin{abstract}
Agentic large language models are increasingly used in digital forensic analysis, yet their ability to infer relational structure inside undocumented mobile application databases remains poorly understood. In forensic contexts, structurally incorrect inferences can yield results that appear plausible while remaining evidentially unsound. This work evaluates agentic structural inference as an isolated capability, treating execution success and structural correctness as distinct evaluation axes. It examines how an agent reconstructs table relationships, linking attributes, and executable join paths when given only a raw database and a natural-language investigative prompt.

We apply a fixed, deterministic evaluation pipeline to two contrasting SQLite repositories: Android's SMS database with stable identifier propagation, and Snapchat's database with irregular schemas, ephemeral identifiers, and polymorphic relationships. Using expert-verified SQL ground truth, we evaluate (i) structural correctness of inferred relational links, (ii) execution coherence under multi-table reasoning, and (iii) robustness and failure modes of inferred structure when execution succeeds but relational interpretation diverges from expert ground truth. Evaluation is performed independently of semantic interpretation, with full queries and execution traces provided in the Appendix.

Results show that structural inference remains reliable in regular schemas but degrades sharply as schema ambiguity increases, frequently producing structurally plausible yet incorrect joins that execute successfully. These findings clarify where schema-agnostic agentic reasoning can support forensic analysis, how its robustness degrades under realistic schema irregularities, and why additional verification remains essential before inferred relationships can be treated as reliable evidence.
\end{abstract}

\section{Introduction}

Mobile application databases record communication events, interaction traces, and operational metadata that frequently serve as primary evidence during digital forensic reconstruction \cite{quick2018data,nemetz2018standardized}. In modern mobile systems, these artefacts are commonly stored in \textit{SQLite} repositories whose internal structure evolves rapidly. Application vendors revise schemas, repurpose fields, and modify relational design across versions, often without documentation. As a result, many real world databases are only partially understood at the time of analysis \cite{garfinkel2010digital,freiling2011androidSmartphones}. Even routine forensic queries therefore require analysts to infer table roles, linking fields, and implicit join paths through manual inspection and iterative querying \cite{case2019practical}.

Most existing forensic extraction approaches assume structural stability. Template driven parsers and schema specific tools operate correctly only when database layouts conform to predefined expectations \cite{daraghmi2023forc,meng2019bring2lite}. When naming conventions drift or when linking attributes become indirect or weakly propagated, these approaches degrade or fail. Manual reconstruction remains possible but is slow and inconsistent, particularly in repositories where identifiers are unstable or distributed across auxiliary tables \cite{goncalves2022datasetGap}. Recent studies show that large language models can generate SQL and assist with schema interpretation under controlled conditions \cite{hong2024surveyTextToSQL,painter2024automating}. However, their behaviour in forensic contexts where schemas are undocumented, relational cues are ambiguous, and evidential correctness is critical remains insufficiently examined \cite{roman2025languageModelsForensics,alzubaidi2024llmforensics}.

This work isolates a narrower and more fundamental problem: how an \textit{agentic} reasoning system infers relational structure inside an undocumented mobile application database when provided only with the raw repository and a natural language investigative prompt. Rather than evaluating semantic retrieval or message interpretation, we focus on the structural inference stage itself. Specifically, we examine how the agent identifies relevant tables, selects linking attributes, proposes join paths, and constructs executable SQL that encodes these structural hypotheses. A key distinction in forensic settings is between queries that merely execute and queries whose inferred structure matches expert validated relational pathways. We therefore evaluate both execution behaviour and the robustness of inferred structure under schema ambiguity.

To observe this behaviour under controlled yet realistic conditions, we evaluate a fixed agent configuration across the same set of investigative tasks on two real mobile repositories selected to represent contrasting schema design characteristics. Android's \texttt{mmssms.db} represents a regular schema with stable identifiers and consistent propagation, while Snapchat's \texttt{main.db} exhibits irregular naming, polymorphic tables, and ephemeral relational links. These repositories are not treated as application specific case studies, but as contrasting structural environments that expose how schema regularity influences inference stability. All prompts, expert validated ground truth queries, execution results, and validation traces are provided in a repository referenced in the Appendix to support reproducibility and independent inspection.

\subsection*{Novelty and Contributions}

This work makes the following contributions:

\begin{itemize}
\item A focused empirical study of agentic structural inference in undocumented mobile databases that explicitly separates relational reasoning from semantic retrieval, and establishes execution success and structural correctness as distinct evaluation axes so that structurally incorrect but executable queries are surfaced rather than masked.
\item A reproducible evaluation framework that applies identical prompts, agent configuration, execution constraints, and scoring criteria across heterogeneous schema designs, supporting controlled comparison between a regular schema (\texttt{mmssms.db}) and an irregular, polymorphic schema (\texttt{main.db}).
\item Characterisation of agent behaviour across investigative tasks involving join path construction, linking attribute inference, and structural filtering, with complete SQL traces available for inspection.
\item Identification of recurring failure modes under schema ambiguity, including key mismatch, join path drift, auxiliary table interference, and structural overconstraint, with discussion of their implications for evidential reliability.
\end{itemize}

\subsection*{Research Questions}

\textbf{RQ1:} To what extent can an agentic system correctly infer the relational structure required to answer forensic queries when operating on undocumented mobile application databases?

\textbf{RQ2:} How stable are the agent's inferred relational structures when translated into executable multi table SQL, particularly for tasks involving implicit keys or multi stage relational reconstruction?

\textbf{RQ3:} When queries execute successfully, how often does the inferred relational structure still diverge from expert ground truth, and what failure modes explain these divergences?

These research questions evaluate whether schema agnostic agentic reasoning can reconstruct relational pathways from structural cues alone, whether inferred structure survives execution, and whether successful execution can still mask structurally plausible but incorrect joins that introduce evidential risk.

\section{Related Work}
\label{sec:related}
Research in mobile database forensics spans multiple areas relevant to this study, including parser based extraction, schema variability, structural ambiguity, and automated reasoning over structured data. While each line of work addresses aspects of forensic analysis, none directly evaluates how an agent infers, executes, and validates relational structure inside undocumented mobile application databases under controlled and reproducible execution constraints.

Parser based extraction has long formed the foundation of forensic database analysis. These systems rely on predefined schema templates that encode expected table layouts, field semantics, and join relationships. Quick and Choo demonstrated that such approaches depend heavily on stable data subsets and degrade as application schemas evolve \cite{quick2018data}. Nemetz et al. further showed that even minor schema changes can break compatibility with widely used \textit{SQLite} extraction tools \cite{nemetz2018standardized}. Automated systems such as FORC improve scalability for Android database analysis but remain dependent on application specific schema knowledge \cite{daraghmi2023forc}. Freiling and colleagues highlighted that frequent schema evolution across mobile platforms undermines deterministic parsing and forces analysts to rely on manual reconstruction \cite{freiling2011androidSmartphones}. These approaches are effective when relational intent is explicit, but they neither infer implicit links nor adapt to undocumented schemas.

Work on dataset diversity emphasizes the gap between publicly available corpora and real world mobile application databases. Gon\c{c}alves et al. documented substantial structural divergence across mobile repositories, demonstrating that schema design varies widely even within the same application category \cite{goncalves2022datasetGap}. Synthetic and standardized datasets improve experimental control and reproducibility, yet still require manual identification of relational patterns and linking attributes \cite{gobel2023datasetSynthesisMobile,nemetz2018standardized}. While these efforts support benchmarking, they do not provide mechanisms for evaluating whether automated systems can correctly infer relational structure in the absence of schema documentation.

Studies of structural ambiguity focus on how incomplete, inconsistent, or corrupted \textit{SQLite} states complicate forensic interpretation. Meng et al. showed that damaged pages and partial records often require investigators to reconstruct schema knowledge manually \cite{meng2019bring2lite}. Case demonstrated that irregular naming conventions and weak relational cues hinder systematic reasoning about table roles and join paths \cite{case2019practical}. These studies address low level irregularities in storage and recovery, but they do not examine whether relational structure can be inferred, executed, and validated automatically when schema cues are incomplete or misleading.

Recent work has explored the use of large language models for structured query generation and schema interpretation. Surveys report strong performance when schemas are well defined, but significant degradation when field names are ambiguous or foreign key relationships are absent \cite{hong2024surveyTextToSQL}. Painter et al. demonstrated that contextual information can improve SQL generation, though their approach assumes relatively stable relational patterns \cite{painter2024automating}. Adjacent work on machine-learning-based anomaly detection in structured records, including efficient detection methods developed for IoT and healthcare data streams \cite{desai2025shield}, illustrates that learned models can flag structurally inconsistent entries, although such methods typically presume known schema layouts and do not generalize to undocumented relational structure. Forensic-focused analyses caution that language models may infer undocumented relationships without sufficient evidential justification, raising concerns about reliability and reproducibility in investigative contexts \cite{roman2025languageModelsForensics,alzubaidi2024llmforensics}. However, these studies typically evaluate syntactic correctness or query executability, rather than separating execution success from structural correctness or analysing failure modes when execution succeeds but inferred structure is wrong.

Taken together, prior work addresses schema stability, dataset diversity, structural ambiguity, and automated reasoning largely in isolation. No existing study systematically evaluates how relational structure is inferred, how inferred structure survives execution, and how robustness degrades under schema ambiguity in undocumented mobile databases. This study addresses that gap through a controlled empirical evaluation of agentic, schema-agnostic structural inference with deterministic execution, expert-verified scoring, and explicit failure-mode analysis.

\section{Framework and Methodology}
\label{sec:framework}
This study evaluates how an agentic reasoning system infers relational structure inside undocumented mobile application databases. The framework isolates and measures structural inference behaviour under controlled forensic constraints, treating it as a first-class reasoning capability that can be evaluated independently of semantic interpretation, application-specific knowledge, or predefined parsers. A database schema is considered \textit{undocumented} when table relationships, linking attributes, and ordering semantics are not explicitly defined through foreign keys, constraints, or documentation, requiring relational structure to be reconstructed solely from schema-visible signals. Given a raw \textit{SQLite} repository and a natural-language investigative question, the agent infers relational hypotheses, constructs executable SQL, and produces verifiable outputs through a controlled execution process that distinguishes queries that merely execute from queries whose inferred structure aligns with expert-validated relational pathways.

\noindent\textbf{Definition of Structural Inference.}
Structural inference is the hypothesis construction step in which candidate tables, linking attributes, join paths, and ordering constraints are derived from observable structural signals such as identifier reuse across tables, column naming and prefix alignment, timestamp co-occurrence patterns, and similarities in table geometry. No templates, foreign keys, documentation, domain semantics, or external assumptions are permitted. All SQL generated by the agent must be justified by these inferred hypotheses, making structural errors observable rather than implicit.

\subsection{Terminology and Agent Configuration}

Throughout this section, the term \textit{agent} refers to the complete controlled reasoning loop, including structural inference, validation, and execution. The term \textit{model} refers exclusively to the underlying large language model used as a bounded reasoning component. \textit{Linking attributes} are inferred fields that connect records across tables, such as thread identifiers, conversation identifiers, sender or receiver fields, or foreign key like values. A \textit{join path} is the ordered sequence of inferred relationships used to construct executable SQL. \textit{Execution coherence} indicates that a generated query executes without structural contradiction, including invalid joins, unresolved attributes, or empty relational paths.

The agent uses OpenAI GPT-4.1, January 2025 release, with deterministic inference settings. Temperature is fixed at 0.0 and top\_p at 1.0, with fixed token budgets of 4,096 tokens for both input and output. This configuration is identical across all investigative tasks and both databases. Each investigative question triggers exactly one reasoning cycle consisting of schema snapshot acquisition, structural hypothesis formation, SQL generation, and supervised execution. No retries, multi seed sampling, prompt adaptation, or fine tuning are employed. All prompts, parameters, and implementation details are provided in the Appendix.

\subsection{Controlled Execution Pipeline}

The agent operates through a fixed multi-stage pipeline that enforces strict separation between reasoning, validation, and execution. \textbf{Schema snapshot acquisition} obtains a complete enumeration of tables, columns, and declared data types via standard \textit{SQLite} introspection in a strictly read-only manner to preserve evidential integrity. \textbf{Structural hypothesis formation} uses the investigative question and schema snapshot to infer candidate tables, linking attributes, join paths, and ordering constraints based exclusively on schema-visible signals. \textbf{Query formulation} translates the inferred hypothesis into exactly one \texttt{SELECT} or \texttt{WITH} query that encodes the proposed relational structure and filtering logic. \textbf{Schema-aware validation and supervised execution} validate the generated query against the schema snapshot for unknown tables, unknown columns, and structural inconsistencies, then execute valid queries in a sandboxed read-only environment, producing result sets and execution metadata. By construction, this pipeline allows structurally incorrect hypotheses to execute successfully while still diverging from expert ground truth, enabling systematic analysis of robustness and failure modes under schema ambiguity.

\begin{figure}[t]
\centering
\includegraphics[width=\linewidth]{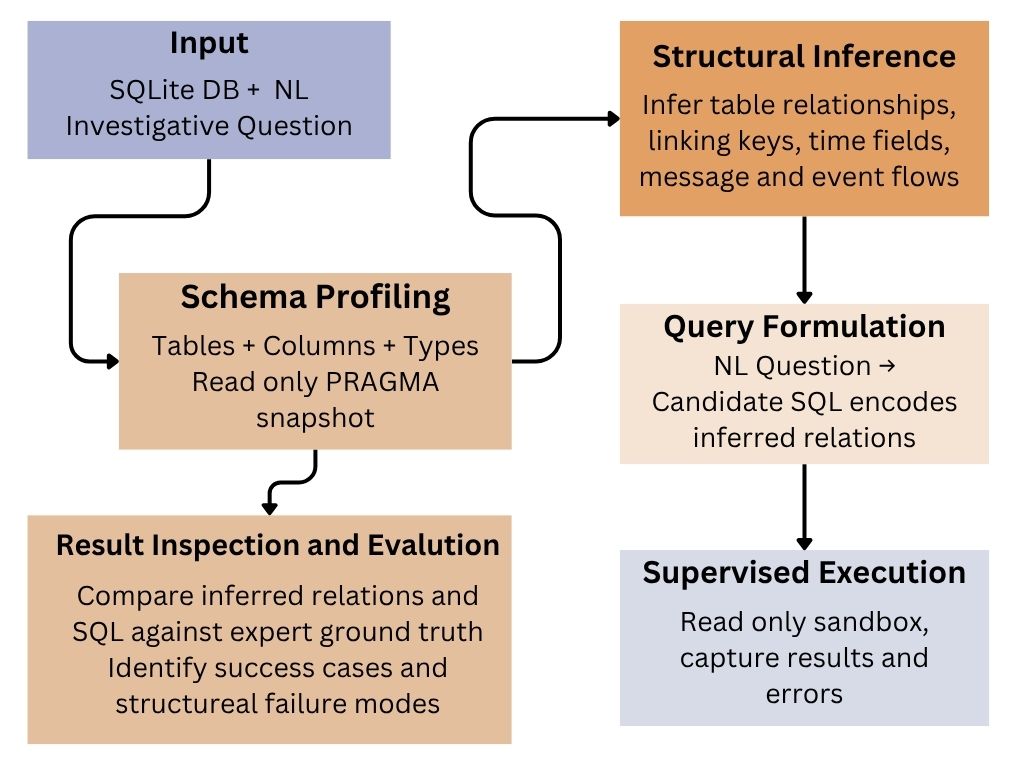}
\caption{Overview of schema profiling, structural inference, SQL formulation, supervised execution, and evaluation.}
\label{fig:workflow}
\end{figure}

\subsection{Structural Inference}

In the absence of foreign keys or documentation, the agent relies on structural cues such as repeated identifiers across tables, substring overlaps, column prefix alignment, timestamp co-occurrence patterns, and parallel table geometry. Prior work shows that many mobile applications rely on implicit relational structure rather than explicit constraints \cite{meng2019bring2lite}. The resulting hypothesis makes all relational assumptions explicit and testable, allowing incorrect inference to manifest as execution divergence or disagreement with expert ground truth.

\begin{figure}[t]
\centering
\includegraphics[width=\linewidth,height=0.40\textheight,keepaspectratio]{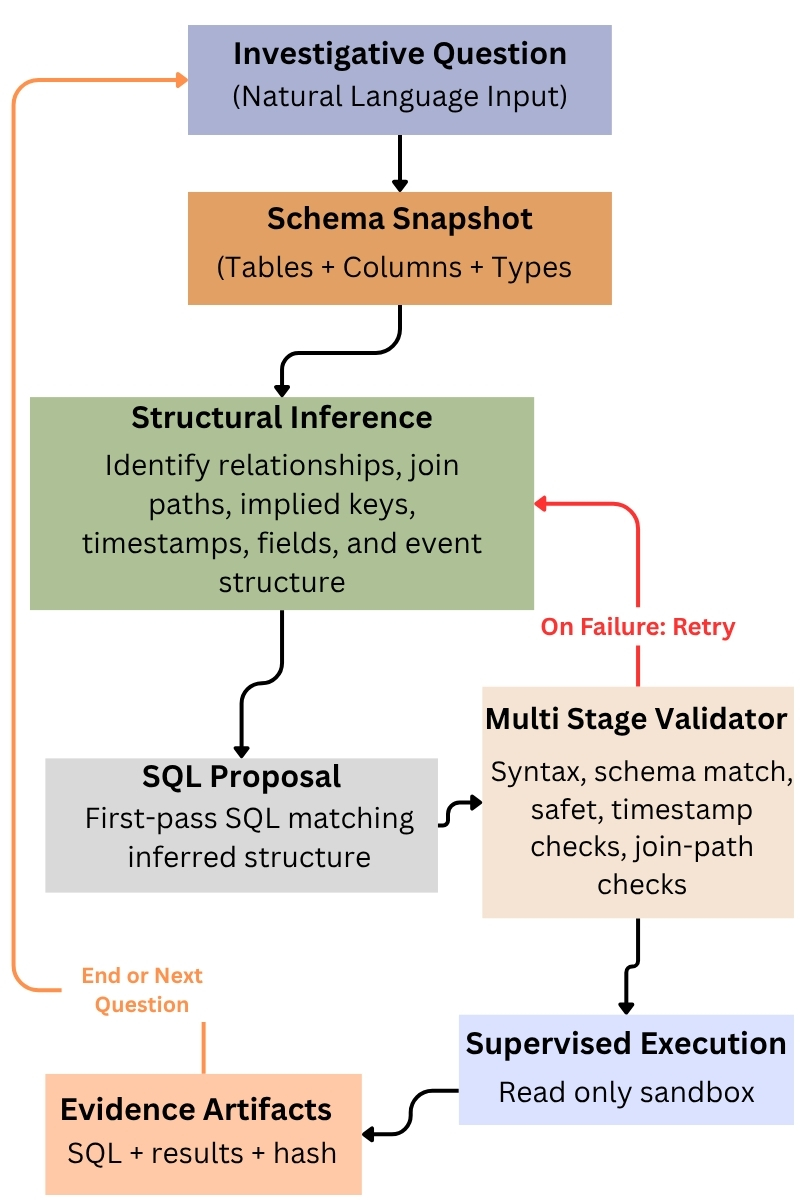}
\caption{Internal reasoning loop showing schema snapshot acquisition, structural hypothesis formation, SQL proposal, validation, and supervised execution.}
\label{fig:internal_loop}
\end{figure}

\subsection{Query Formulation and Supervised Execution}

Structural hypotheses are converted into SQL under strict constraints. Only schema defined tables and columns may be referenced, exactly one query must be produced, and the inferred join path must be followed. A \texttt{LIMIT} clause is added if none is present. Queries that pass structural validation execute in a sandboxed, read only environment. Execution outputs, validation metadata, and error traces are recorded and included in the Appendix.

\subsection{Evaluation Alignment}

The framework is evaluated using the same set of twelve investigative questions across all databases. These questions are designed to probe structural inference across increasing relational complexity, including implicit key recovery and multi table reconstruction. The evaluation aligns framework stages with the research questions as follows: structural hypothesis formation addresses RQ1, supervised execution addresses RQ2, and divergence between executed structure and expert validated ground truth under successful execution supports RQ3. Metrics and scoring procedures are detailed in Section~\ref{sec:empirical_study}.

\section{Empirical Study}
\label{sec:empirical_study}

This section describes the datasets, ground truth construction, and scoring methodology used to evaluate how the agent infers relational structure from undocumented mobile application databases. The evaluation targets structural reasoning only, including the identification of linking attributes, construction of join paths, and preservation of structurally relevant ordering and filtering constraints during execution. Interpretation of message content or semantic relevance is explicitly excluded. All results are quantified using Precision, Recall, and F1 scores computed over projected evidentiary units.

\subsection{Evaluation Datasets}

The study uses two mobile repositories selected to represent contrasting levels of structural stability commonly encountered in forensic analysis. The database \textit{mmssms.db} represents the regular case. It is a compact SMS and MMS store with explicit identifiers, stable thread semantics, and limited schema drift across Android versions. Its predictable structure provides a controlled environment for observing agent behaviour when relational intent is clearly encoded.

Snapchat's \texttt{main.db} represents the irregular case. It contains numerous interdependent tables, ephemeral identifiers, polymorphic record layouts, and inconsistent naming conventions. Relationships are often indirect, weakly propagated, or distributed across auxiliary tables. These properties reflect real world scenarios in which application schemas evolve rapidly without preserving coherent relational structure or documentation.

Together, these repositories form a controlled experimental contrast that forces schema agnostic inference. Across both databases, the agent must reason over implicit links, weak or missing key constraints, overlapping identifiers, and inconsistent timestamp representations. Complete schema snapshots for both repositories are provided in the Appendix.

\subsection{Establishing Ground Truth}

Ground truth was constructed independently for each of the twelve investigative tasks. Two mobile forensics analysts, each with more than five years of professional experience, examined each database to determine the structurally correct interpretation required by each task, identifying the correct join path implied by the prompt, the linking attributes defining cross-table relationships, the relational scope and constraints required to satisfy the query, and the complete row set that a structurally correct query should return. Disagreements were resolved through stepwise verification of identifier propagation, join consistency, and ordering constraints until consensus was reached. This process yields a single expert-validated ground truth query per task. By anchoring evaluation to explicit SQL and row-level outputs, the methodology isolates structural correctness from content variability. All ground truth SQL queries are included in the Appendix.

\subsection{Evaluation Metrics}
\label{sec:metrics}

The evaluation targets structural correctness only. Natural language explanations and intermediate reasoning traces are not considered during scoring. For each database, the twelve investigative questions are grouped into three difficulty tiers: G1 for basic retrieval tasks that primarily assess structural identification, G2 for contextual filtering tasks that require limited relational chaining, and G3 for multi stage or adversarial tasks that stress execution stability and robustness under schema ambiguity. These tiers collectively support analysis of RQ1 through RQ3.

Each question $q$ is associated with a ground-truth query $Q^{\mathrm{GT}}_q$, an agent-generated query $Q^{\mathrm{AG}}_q$, and a key column set $K_q$ representing the evidentiary units of interest. Both queries are executed directly against the original database. Let $R^{\mathrm{GT}}_q$ and $R^{\mathrm{AG}}_q$ denote the resulting row sets. To ensure comparison reflects structural agreement rather than extraneous attributes, both result sets are projected onto the key columns:

\[
G_q = \{\pi_{K_q}(r) \mid r \in R^{\mathrm{GT}}_q\}, \qquad
P_q = \{\pi_{K_q}(r) \mid r \in R^{\mathrm{AG}}_q\}.
\]

All agreement metrics are computed over these projected sets. Computing agreement on key column projections rather than full row sets isolates structural alignment on the evidentiary units of interest. As a consequence, two queries returning structurally distinct but evidentially equivalent rows may receive different scores when their projections disagree, a property revisited in the validity discussion of Section~\ref{sec:limitations}.

\subsubsection{Row Level Agreement Metrics}

For each question $q$, Precision and Recall are defined as:

\[
\mathrm{Precision}_q =
\frac{|P_q \cap G_q|}{|P_q|}
\quad\text{with value 0 if } |P_q| = 0,
\]

\[
\mathrm{Recall}_q =
\frac{|P_q \cap G_q|}{|G_q|}
\quad\text{with value 0 if } |G_q| = 0.
\]

The F1 score is computed as the harmonic mean of Precision and Recall:

\[
\mathrm{F1}_q =
2 \cdot \frac{\mathrm{Precision}_q \cdot \mathrm{Recall}_q}
{\mathrm{Precision}_q + \mathrm{Recall}_q}.
\]

False positives, defined as $P_q \setminus G_q$, indicate spurious relationships introduced by incorrect join paths or filtering logic. False negatives, defined as $G_q \setminus P_q$, indicate structurally valid relationships that the agent failed to recover. Temporal correctness is evaluated implicitly through row-level agreement after projection onto expert-defined key columns, rather than through a separate temporal distance or alignment metric.

\paragraph{Edge Cases.} Four boundary conditions warrant explicit treatment so that low scores in Table~\ref{tab:rq1_full} are interpreted consistently. First, when both $P_q$ and $G_q$ are empty, the result is treated as a structurally valid agreement because the absence condition is satisfied by both queries. Second, when $G_q$ is non-empty but $P_q$ is empty, Recall is set to zero, reflecting complete failure to recover the expected evidentiary set. Third, when $P_q$ is non-empty but $G_q$ is empty, Precision is set to zero, reflecting introduction of spurious records where none should exist. Fourth, when $|P_q \cap G_q| = 0$ despite both sets being non-empty, both Precision and Recall evaluate to zero. This last case accounts for the 0.00 entries observed in several Snapchat tasks, which indicate complete misalignment of inferred relational structure rather than execution failure.

\paragraph{Illustrative Example.}
For Question 7 applied to Snapchat's \texttt{main.db}, the expert-validated ground truth identifies three text-bearing records, $G_q = \{408, 409, 450\}$, while the agent retrieves these records along with two unrelated rows from the table \texttt{LocalConversationInteraction}, giving $P_q = \{1, 2, 408, 409, 450\}$. The intersection is $P_q \cap G_q = \{408, 409, 450\}$, so Precision is $3/5 = 0.60$ and Recall is $3/3 = 1.00$, yielding an F1 score of $0.75$. This example illustrates a recurring failure mode in irregular schemas, where the agent produces joins that are structurally admissible and executable but diverge from expert-validated relational structure. Additional per-task outputs and extended examples are provided in Section~\ref{sec:appendix}.

\subsection{RQ1: How well does an agentic system recover structural links in undocumented mobile databases?}

RQ1 evaluates the central question of this study: whether an agentic system can recover the relational structure required to answer forensic queries when operating on undocumented mobile application databases. Unlike prior Text-to-SQL systems or parser-driven approaches, this evaluation requires the agent to infer structure strictly from schema-visible evidence. No foreign keys, naming templates, domain semantics, or application-specific assumptions are provided. The twelve investigative tasks are organized into three inference tiers. G1 tasks require single-hop retrieval from a primary table. G2 tasks require two-hop linking across related tables. G3 tasks introduce multi-hop dependencies, malformed or weak identifiers, timestamp drift, or absence-based retrieval conditions. This tiered design isolates structural inference as an independent capability rather than conflating it with semantic reasoning.

For each task, the agent is required to infer appropriate linking attributes, construct a valid join path, and produce an executable query. Evaluation is performed using two complementary measures. First, the inferred structural keys and join structure are assessed implicitly through row-level agreement with expert-validated ground truth. Second, Precision, Recall, and F1 scores are computed over projected evidentiary units as defined in Section~\ref{sec:metrics}. Together, these measures capture whether the agent selects correct relational pathways and whether those selections recover the intended evidentiary records. Table~\ref{tab:rq1_full} reports all values, with results for SMS and Snapchat evaluated separately.

\begin{table*}[!htbp]
\centering
\caption{RQ1: Structural Keys and Row-Level Agreement Metrics Across Both Databases}
\label{tab:rq1_full}
\renewcommand{\arraystretch}{1.18}
\begin{tabular}{|p{10.2cm}|c|c|c|c|c|}
\hline
\textbf{Question} &
\textbf{Database} &
\textbf{Structural Key} &
\textbf{Precision} &
\textbf{Recall} &
\textbf{F1} \\
\hline

\multirow{2}{ 10.2cm}{\textbf{Q1.} List all messages stored in the database.}
& SMS      & message\_id     & 1.00 & 1.00 & 1.00 \\ \cline{2-6}
& Snapchat & conversation\_id & 0.40 & 0.67 & 0.50 \\ \hline

\multirow{2}{ 10.2cm}{\textbf{Q2.} Show all unique phone numbers that appear in the message records.}
& SMS      & address     & 0.85 & 1.00 & 0.92 \\ \cline{2-6}
& Snapchat & record\_id  & 1.00 & 0.07 & 0.13 \\ \hline

\multirow{2}{ 10.2cm}{\textbf{Q3.} List the most recent messages.}
& SMS      & message\_id     & 1.00 & 1.00 & 1.00 \\ \cline{2-6}
& Snapchat & identifier       & 0.00 & 0.00 & 0.00 \\ \hline

\multirow{2}{ 10.2cm}{\textbf{Q4.} Show all unread messages.}
& SMS      & message\_id     & 1.00 & 1.00 & 1.00 \\ \cline{2-6}
& Snapchat & media\_value    & 0.86 & 0.24 & 0.38 \\ \hline

\multirow{2}{ 10.2cm}{\textbf{Q5.} Retrieve all messages exchanged with a specific phone number starting with +1.}
& SMS      & message\_id     & 1.00 & 1.00 & 1.00 \\ \cline{2-6}
& Snapchat & conversation\_id & 0.00 & 0.00 & 0.00 \\ \hline

\multirow{2}{ 10.2cm}{\textbf{Q6.} Show all messages containing possible OTP or verification patterns.}
& SMS      & message\_id     & 0.88 & 0.54 & 0.67 \\ \cline{2-6}
& Snapchat & identifier       & 1.00 & 0.04 & 0.08 \\ \hline

\multirow{2}{ 10.2cm}{\textbf{Q7.} Group all messages by thread, count messages per thread, and list their phone number.}
& SMS      & thread\_id    & 0.82 & 1.00 & 0.90 \\ \cline{2-6}
& Snapchat & record\_id    & 0.60 & 1.00 & 0.75 \\ \hline

\multirow{2}{ 10.2cm}{\textbf{Q8.} List all MMS records or entries missing or lacking a body field.}
& SMS      & message\_id  & 0.00 & 0.00 & 0.00 \\ \cline{2-6}
& Snapchat & record\_id  & 0.98 & 0.55 & 0.70 \\ \hline

\multirow{2}{ 10.2cm}{\textbf{Q9.} Find all messages exchanged on 29 January 2020.}
& SMS      & message\_id  & 1.00 & 1.00 & 1.00 \\ \cline{2-6}
& Snapchat & record\_id  & 0.00 & 0.00 & 0.00 \\ \hline

\multirow{2}{ 10.2cm}{\textbf{Q10.} Identify periods of unusually high messaging activity and list associated phone numbers.}
& SMS      & hour\_bucket  & 0.00 & 0.00 & 0.00 \\ \cline{2-6}
& Snapchat & conversationId & 0.00 & 0.00 & 0.00 \\ \hline

\multirow{2}{ 10.2cm}{\textbf{Q11.} Identify message entries with structurally inconsistent fields (missing timestamps, invalid thread identifiers, mismatched metadata).}
& SMS      & message\_id  & 0.00 & 0.00 & 0.00 \\ \cline{2-6}
& Snapchat & record\_id  & 0.01 & 0.17 & 0.02 \\ \hline

\multirow{2}{ 10.2cm}{\textbf{Q12.} Detect structurally anomalous message records with schema-level inconsistencies (null timestamps, negative IDs, mismatched relations).}
& SMS      & message\_id  & 0.00 & 0.00 & 0.00 \\ \cline{2-6}
& Snapchat & record\_id  & 0.00 & 0.00 & 0.00 \\ \hline

\end{tabular}

\vspace{1mm}
\footnotesize
Scores reported separately for SMS (\textit{mmssms.db}) and Snapchat (\textit{main.db}).
\end{table*}

Results reveal a clear and consistent pattern. In \textit{mmssms.db}, the agent reliably recovers the intended structural links for all G1 and G2 tasks. Explicit identifiers, stable propagation of message and thread relationships, and uniform timestamp semantics support consistent structural inference, reflected in near-perfect agreement scores. Failures emerge primarily in G3 tasks that require multi-hop reasoning or depend on absence-based conditions, where structural cues are inherently weaker.

In contrast, performance on Snapchat's \texttt{main.db} degrades as soon as cross-table inference is required. Weakly propagated UUIDs, inconsistent identifier fields, and structurally similar tables lead to systematic key mismatches, such as confusion between \texttt{conversation\_id} and \texttt{container\_id}. These mismatches directly correspond to increases in both false positives and false negatives, as reflected in declining Precision, Recall, and F1 scores. As relational distance increases across G2 and G3 tasks, row-level agreement deteriorates accordingly.

Taken together, the results for RQ1 demonstrate that recovery of structural links is strongly conditioned on schema regularity. When identifiers are explicit and consistently propagated, schema-agnostic inference can accurately reconstruct relational pathways. When identifiers are ephemeral, polymorphic, or weakly aligned across tables, recovery of correct structural links becomes unreliable. These findings motivate the subsequent analysis of execution stability and robustness in RQ2 and RQ3.

\subsection{RQ2: How stable are the agent's structural inferences when retrieval requires multi-table reasoning, implicit key detection, or temporal reconstruction?}

RQ2 evaluates whether the relational hypotheses identified in RQ1 remain stable when translated into executable SQL. Stability in this context is operational rather than semantic. A structural hypothesis is considered stable if the SQL query produced from it executes successfully against the datastore without introducing structural contradictions. Such contradictions include invalid joins, missing linkage attributes, unintended Cartesian products, empty relational paths, or schema-level execution errors. We refer to this property as \emph{execution coherence}.

Execution coherence is evaluated as a binary outcome for each task. A query is deemed coherent if it passes schema-aware validation and executes successfully in the sandboxed environment defined in Section~\ref{sec:framework}. Queries that fail validation, reference incompatible join paths, or produce structurally contradictory results are deemed incoherent. Importantly, execution coherence does not assess whether the resulting rows are correct with respect to ground truth; that assessment is reserved for RQ1 and RQ3. RQ2 isolates whether inferred structure survives operational realization at all.

Table~\ref{tab:rq2_final} summarizes execution coherence across task categories and datastores. Tasks are grouped according to the same G1 through G3 difficulty tiers defined earlier. Because the benchmark isolates structural inference rather than semantic interpretation, coherence failures directly expose breakdowns in schema-agnostic structural reasoning, including weak identifier propagation, implicit key mismatch, multi-table ambiguity, and incompatible temporal representations.

\begin{table}[htbp]
\centering
\caption{Execution coherence across evaluation categories.}
\renewcommand{\arraystretch}{1.12}
\begin{tabular}{|c|l|c|l|}
\hline
\textbf{Database} &
\textbf{Category} &
\begin{tabular}[c]{@{}c@{}}\textbf{Coherent /}\\\textbf{Executions}\end{tabular} &
\textbf{Failure Modes} \\ \hline

\multirow{3}{*}{mmssms.db}
 & Basic Retrieval        & 4/4 & None \\ \cline{2-4}
 & Filtering and Context  & 3/4 & Filter miss \\ \cline{2-4}
 & Advanced Reasoning     & 2/4 & Join drift \\ \hline

\multirow{3}{*}{main.db}
 & Basic Retrieval        & 2/4 & Key mismatch \\ \cline{2-4}
 & Filtering and Context  & 3/4 & Attribute drop \\ \cline{2-4}
 & Advanced Reasoning     & 1/4 & Path break \\ \hline

\end{tabular}

\vspace{2mm}
\footnotesize
Execution coherence indicates whether generated SQL executes without invalid joins, missing linkage attributes, or structurally contradictory execution paths. SMS refers to \textit{mmssms.db}. Snapchat refers to \textit{main.db}.
\label{tab:rq2_final}
\end{table}

Results for the SMS repository (\textit{mmssms.db}) indicate high execution coherence across G1 and G2 tasks. Explicit thread identifiers and consistent timestamp representations allow most inferred join paths to execute without contradiction. Coherence failures are concentrated in G3 tasks that rely on absence-based retrieval or malformed record detection, where minor misalignment in linking attribute selection leads to join drift or incomplete relational paths.

In contrast, Snapchat's \texttt{main.db} exhibits substantially lower execution coherence once tasks require implicit key recovery or multi-table reconstruction. Ephemeral UUIDs do not propagate consistently across tables, timestamp representations vary across event streams, and multiple tables expose structurally similar columns encoding unrelated relationships. These conditions lead to attribute loss, incompatible joins, or complete join path failure during execution.

Overall, RQ2 demonstrates that the stability of agent-inferred structure during execution is strongly dependent on schema regularity and identifier propagation. Regular schemas preserve inferred structure at the point of realization, while irregular or polymorphic schemas frequently cause structurally valid-looking hypotheses to fail when instantiated as executable SQL. This motivates the subsequent robustness analysis in RQ3, which examines how inferred structure behaves when execution succeeds but relational interpretation diverges from expert ground truth.

\subsection{RQ3: How robust is structural inference when execution succeeds but relational interpretation diverges from ground truth?}
\label{sec:rq3_results}

RQ3 examines the robustness of agentic structural inference under conditions where query execution succeeds but the inferred relational structure does not fully align with expert-validated ground truth. Unlike RQ2, which evaluates whether inferred structure can be operationalized without contradiction, RQ3 focuses on how inference quality degrades \emph{after successful execution}. This distinction isolates robustness from executability.

Robustness is evaluated only over queries that satisfy execution coherence as defined in RQ2. Within this subset, we analyze how inferred linking attributes, join paths, and relational alignment compare to expert ground truth using the row-level Precision, Recall, and F1 metrics defined in Section~\ref{sec:metrics}. Divergence in this setting reflects structurally admissible but evidentially incorrect interpretations rather than execution failure.

\subsubsection{Effect of Relational Complexity}

For low-complexity tasks involving single-table access or direct attribute filtering (G1), robustness is consistently high across both repositories, and successful execution almost always coincides with correct structural inference. As relational complexity increases (G2), robustness becomes sensitive to identifier propagation and schema regularity: \textit{mmssms.db} retains alignment with ground truth through consistent thread-identifier reuse and uniform table geometry, whereas Snapchat's \texttt{main.db} exhibits frequent divergence as multiple candidate linking attributes satisfy structural similarity constraints and lead the agent toward alternative but plausible relational pathways. The most pronounced degradation occurs in G3 tasks requiring multi-stage joins, implicit key recovery, or cross-table temporal alignment, where several structurally valid interpretations may exist simultaneously and the agent may infer a join path that is internally coherent and executable but differs from the expert-selected relational structure. These outcomes reflect intrinsic schema ambiguity rather than loss of schema awareness or execution instability.

\subsubsection{Platform-Level Robustness Differences}

Systematic robustness differences are observed between the two repositories. Android's \textit{mmssms.db} maintains relatively stable robustness across increasing relational depth due to explicit identifier propagation and limited schema polymorphism, so when execution succeeds, inferred structure typically remains close to expert ground truth. Snapchat's \texttt{main.db} exhibits substantially lower robustness under comparable conditions because polymorphic tables, auxiliary caches, and loosely coupled identifiers increase the number of structurally admissible interpretations; successful execution does not reliably imply correct relational reconstruction, and divergence is attributable to structural ambiguity rather than syntactic or validation failure.

\subsubsection{Observed Robustness Failure Modes}

Among successfully executed queries that diverge from expert ground truth, four recurring structural failure modes are observed: \textbf{ambiguous linking attribute selection}, where multiple fields satisfy schema-level similarity cues and lead to selection of an alternative but incorrect join key; \textbf{join path substitution}, where an executable join path differs from the expert-validated relational pathway while remaining structurally coherent; \textbf{auxiliary table interference}, where metadata or cache tables introduce structurally valid but evidentially irrelevant relationships that compete with primary joins; and \textbf{structural overconstraint}, where inferred filters or grouping constraints preserve executability but eliminate relevant records, reducing recall without triggering validation failure. These failure modes occur without invalid joins, schema errors, or execution rejection, and represent degradations in inferred relational structure rather than instability of the execution pipeline.

\subsubsection{Sources of Robustness Failure}

The four failure modes arise from distinct underlying causes. Ambiguous linking attribute selection and auxiliary table interference are predominantly attributable to intrinsic schema ambiguity in undocumented databases, where multiple structural cues compete for the same relational role with no documented constraint to disambiguate. Join path substitution reflects weak identifier propagation combined with the agent's tendency to favour structurally simpler paths under deterministic decoding. Structural overconstraint is most directly attributable to model inference behaviour under the fixed prompting strategy, since alternative prompting could relax inferred filters without altering the database. A residual share is attributable to the evaluation protocol, specifically the dependence on a single expert-validated join path when alternative valid decompositions may exist. Schema-driven failures are intrinsic to the forensic environment and cannot be eliminated by improved reasoning alone, whereas model-driven and protocol-driven failures define directions for future system and benchmark refinement.

\subsubsection{Summary}

RQ3 demonstrates that agentic structural inference degrades gracefully under increasing schema ambiguity. When divergence occurs, failures are bounded, systematic, and attributable to intrinsic properties of undocumented schemas rather than uncontrolled generation. Successful execution alone is therefore insufficient to establish evidential reliability in irregular databases, motivating explicit robustness and failure-mode analysis in forensic applications.

\section{Discussion}

The results clarify both the capabilities and the limits of agentic structural inference in forensic database analysis. By isolating structural reasoning from semantic interpretation, the evaluation identifies which aspects of undocumented database analysis can be supported reliably by schema-agnostic inference and which remain inherently fragile. Structural inference is most effective when relational intent is encoded through explicit and consistently propagated identifiers: in such settings, inferred join paths remain stable, execution coherence is high, and divergence from expert-validated ground truth is rare, supporting the use of agentic reasoning for early-stage exploration and hypothesis generation. Under schema ambiguity, structural inference degrades in predictable ways, as weak identifier propagation, polymorphic fields, auxiliary table interference, and incompatible timestamp conventions introduce multiple structurally admissible interpretations, and successful execution alone does not guarantee evidential correctness.

These findings have direct implications for forensic practice. Agentic structural inference should be treated as an assistive capability rather than an authoritative source of relational truth. Inferred relationships are most reliable when identifiers are explicit and relational distance is limited, while increasing structural complexity necessitates additional verification, particularly for multi-hop joins and temporal reconstruction tasks. More broadly, this study underscores the importance of evaluating AI-assisted forensic tools at the level of structural reasoning rather than relying solely on query executability or surface-level accuracy.

\subsection{Structural Robustness and Evidential Risk}

The four failure modes characterized in Section~\ref{sec:rq3_results} occur without invalid joins, schema errors, or validation rejection, which means structurally plausible and executable queries may still encode incorrect evidential relationships when schema ambiguity is high, particularly under multi-stage joins, implicit key recovery, or cross-table temporal alignment.

\subsubsection{Operational Recommendations}

Each failure mode admits a corresponding verification action that forensic analysts can apply before treating inferred relationships as evidentiary, and these verifications can be performed within the same controlled execution pipeline used during inference, preserving the read-only and reproducible properties required for forensic analysis. Ambiguous linking attribute selection should be addressed by re-executing the inferred query against alternative candidate keys exhibiting comparable structural similarity and then comparing returned row counts and identifier overlap; substantial divergence between alternative joins indicates that the inferred attribute is not uniquely supported by schema evidence. Join path substitution should be probed by sampling intermediate joins along the inferred chain and verifying identifier propagation at each hop, rather than accepting the end-to-end query as a single unit. Auxiliary table interference should be tested by repeating the query with cache, metadata, or session-state tables explicitly excluded from the join scope, and then comparing result sets to detect contributions from non-evidentiary sources. Structural overconstraint should be probed by selectively relaxing inferred filtering and grouping constraints to determine whether relevant records were excluded by inference rather than by the underlying data. Treated together, these actions translate the failure-mode taxonomy into a concrete verification protocol that practitioners can adopt without retraining or modifying the agent.

\subsubsection{Graduated Trust Model}

The evidence accumulated across RQ1 through RQ3 supports a graduated trust model for agentic structural inference in forensic settings. When schema regularity is high, identifier propagation is consistent, and relational distance is short, inferred structure can be treated as provisionally reliable subject to standard sanity checks. As schema irregularity, relational depth, or identifier ambiguity increase, inferred structure should be treated as a hypothesis requiring corroboration through the verification actions described above, manual schema inspection, or comparison against independent extraction tools. This graduated model preserves the practical value of agentic exploration while protecting against the evidential risks introduced by structurally plausible but incorrect inference, and it provides forensic teams with explicit criteria for deciding when inferred relationships can be relied upon and when they must be re-examined.

\section{Limitations and Threats to Validity}
\label{sec:limitations}

This study deliberately scopes its analysis to structural inference as an isolated reasoning capability. Reported results characterize structural soundness rather than end-to-end forensic accuracy. The validity considerations below follow from this scoping decision and from the controlled experimental design adopted to support reproducibility.

\subsubsection{Single Model Configuration}
The evaluation uses a single underlying model, OpenAI GPT-4.1 (January 2025 release), under a single fixed prompting strategy. This choice was made to maximize reproducibility and isolate structural inference behaviour under controlled conditions. It limits direct generalization across model families, parameter scales, and prompting designs. Comparative evaluation across multiple reasoning systems is a natural extension that the framework is explicitly designed to support, as the pipeline, scoring procedure, and benchmark tasks are all model-agnostic.

\subsubsection{Deterministic Single-Run Protocol}
Inference is performed at temperature 0.0 with fixed decoding parameters, producing one execution per task. This protocol prioritizes reproducibility and per-query attribution of failure modes over variance reporting. The study therefore does not present confidence intervals or stochastic-decoding statistics. Variance characterization under non-deterministic settings is a valid extension that would complement, rather than replace, the deterministic baseline established here, particularly for cross-system comparison where repeated sampling is standard.

\subsubsection{Benchmark Scope}
The evaluation considers two real mobile repositories and twelve investigative tasks organized across three difficulty tiers. The repositories were selected to expose contrasting structural conditions, namely a regular schema with explicit identifiers and an irregular schema with ephemeral and polymorphic identifiers, rather than to represent the full diversity of mobile application schemas. Generalization across additional application categories, schema generations, and forensic task types remains open and would require coordinated benchmark expansion. The small task count per tier should be interpreted as a controlled probe of structural inference behaviour rather than a population-level estimate.

\subsubsection{Expert-Defined Ground Truth and Alternative Valid Queries}
Ground truth was constructed by two senior mobile forensics analysts and reflects one expert-validated relational interpretation per task. In undocumented schemas, alternative join paths may yield structurally distinct queries that nevertheless return evidentially equivalent records. Precision, Recall, and F1 are computed over expert-defined key column projections, so structurally divergent but evidentially equivalent agent outputs may receive lower scores than their forensic value warrants. This is a known property of relational benchmarks under schema ambiguity rather than a measurement defect, but it qualifies the interpretation of low agreement scores in irregular schemas and motivates the inclusion of full SQL traces in the Appendix so that reviewers can inspect alternative interpretations directly.

\subsubsection{Cross-Database Reasoning and Prompt Sensitivity}
Each datastore is analysed independently. The study does not evaluate cross-database structural correlation or multi-source reconstruction, both of which arise frequently in real forensic workflows where evidence is distributed across application stores, system databases, and external corroborating artefacts. Although execution is deterministic and sandboxed, results remain sensitive to prompt formulation. Prompt design was held fixed across all tasks and databases for comparability, and prompt-level optimization was deliberately excluded from the experimental scope to avoid confounding model capability with prompt engineering.

The framework itself is model-agnostic, prompt-agnostic, and benchmark-extensible. The limitations above identify the dimensions along which controlled comparative evaluation can be expanded without altering the core methodology.

\section{Future Work}

Several extensions follow naturally from the current evaluation. Comparative evaluation across multiple model families, parameter scales, and prompting strategies, including non-agentic Text-to-SQL baselines, would clarify how structural inference behaviour varies across reasoning systems and would test whether the failure-mode taxonomy reported here generalizes beyond a single agent configuration. Variance and confidence-interval reporting under stochastic decoding would complement the deterministic protocol used here and enable statistical comparison between systems on the same benchmark. Expanded benchmarking across additional mobile application categories, schema generations, and forensic task types would test the generality of the structural failure modes observed in this study and support population-level claims that the present controlled probe cannot establish. Beyond model and benchmark extensions, cross-database structural correlation would better reflect investigations that align events across multiple local stores, more explicit temporal reasoning may improve stability where timestamps drift or follow incompatible conventions, and evaluating cloud-backed, non-SQLite, and hybrid application artefacts would broaden assessment of generality across storage paradigms. Storage-layer technologies that govern how mobile data is physically written and encoded, including phase-change memory write optimization and encoding strategies \cite{desai2024wire,desai2025smartwrite}, lie outside the schema-agnostic scope of this study but represent complementary directions for end-to-end forensic acquisition pipelines. Integrating lightweight semantic or internal consistency checks could further constrain inference when structural cues are weak, implicit, or inconsistently propagated, without undermining the schema-agnostic focus of the present approach. Collectively, these directions define a research agenda in which the framework introduced here serves as a reproducible foundation for comparative evaluation of agentic forensic reasoning systems.

\section{Conclusion}

This study evaluated how an agentic reasoning system infers relational structure inside undocumented mobile application databases under controlled and reproducible conditions. By isolating structural inference from semantic interpretation, and by treating execution success and structural correctness as distinct evaluation axes, the analysis identifies both the conditions under which schema-agnostic reasoning remains reliable and the points at which it degrades. In regular schemas such as \textit{mmssms.db}, explicit identifiers and consistent propagation support accurate recovery of join paths and stable execution. In irregular repositories such as Snapchat's \texttt{main.db}, weak or polymorphic identifiers introduce ambiguity that leads to join drift, key mismatch, and execution instability. Successful execution alone does not guarantee structural correctness, and the operational recommendations described in Section~V remain necessary before inferred relationships can be treated as evidentiary. By characterizing both success and failure under controlled conditions, this work clarifies the practical boundaries of agentic structural reasoning in mobile database forensics. The framework, benchmark, and failure-mode taxonomy introduced here provide a reproducible foundation that other researchers can extend along the model, prompt, and dataset dimensions identified above, supporting principled comparison of agentic reasoning systems in forensic database analysis.

\vspace{12pt}

\section{Appendix}
\label{sec:appendix}
All code, schema snapshots, prompts, execution traces, and evaluation artefacts used in this study are available in the public repository:
\begin{center}
\small
\url{https://github.com/jeelkhatiwala/IEEE-CompSac}
\end{center}
The repository contains complete executable notebooks for both \textit{mmssms.db} and \textit{Snapchat main.db}, preserving the full agent workflow including schema profiling, JSON plan generation, plan-to-SQL translation, structural validation, and execution outputs. Ground-truth SQL, structural key definitions, and projected result sets are included for every task.
\end{document}